\documentstyle[epsfig,subfigure,twocolumn]{mn}

\newcommand{\goodgap}{\hspace{\subfigtopskip} \hspace{\subfigbottomskip}}

\title[Angular momentum and the $R_{eff}$\,-\,$M_{\star}$ relation in ETGs]
{Angular momentum transfer and the size\,-\,mass relation in early\,-\,type galaxies}
\author[V.F. Cardone et al.]{V.F. Cardone$^{1}$, A. Del Popolo$^{2}$, P. Kroupa$^{3}$\\
$^1$ Dipartimento di Fisica Generale "Amedeo Avogadro" and INFN - Sez. di Torino, 
Via Pietro Giuria 1, 10125 - Torino, Italy \\ $^2$
Dept. of Physics and Astronomy, Catania University, via S. Sofia 78, 95123 - Catania, Italy \\
$^3$ Argelander Institut f\"ur Astronomie, Auf dem H\"ugel 71, D -
53121 Bonn, Germany }

\date{Accepted xxx, Received yyy, in original form zzz}

\begin{document}

\maketitle

\begin{abstract}

Early\,-\,type galaxies (ETGs) define a narrow strip in the
size\,-\,mass plane because of the observed correlation between
the effective radius $R_{eff}$ and the total stellar mass
$M_{\star}$. When expressed in logarithmic units, a linear
relation, $\log{R_{eff}} \propto \gamma \log{M_{\star}}$, is
indeed observationally found, but the slope $\gamma$ deviates from
the canonical $\gamma = 1/2$ value which can be naively predicted
for a spherically symmetric isotropic system. We propose here that
a transfer of angular momentum to the stellar component induces an
anisotropy in the velocity space thus leading to a modified
distribution function (DF). Assuming an Osipkov\,-\,Merritt like
anisotropic DF, we derive an analytic relation between the slope
$\gamma$ of the size\,-\,mass relation and the slope $\alpha$ of
the angular momentum term in the DF. With this simple model, we
are then able to recover the observed $\gamma$ value provided
$\alpha$ is suitably set. It turns out that an anisotropy profile
which is tangential inside $\sim 0.6 r_a$ and radial outside, with
$r_a$ the anisotropy radius, is able to reproduce the observed
size\,-\,mass relation observed for massive ($M_{\star} \ge 2
\times 10^{10} \ h^{-1} \ {\rm M}_{\odot}$) elliptical galaxies.

\end{abstract}


\section{Introduction}

Notwithstanding the quite large range spanned by their
morphological and photometric properties, early\,-\,type galaxies
(hereafter, ETGs) show several interesting correlations among
their colors, luminosities, velocity dispersions, effective radii
and surface brightness (e.g. Baum 1990, Faber \& Jackson 1976,
Kormendy 1977, Djorgovski \& Davis 1987, Dressler et al. 1987,
Bernardi et al. 2003a). As a most famous example, ETGs are known
to populate a tight plane, referred to as the {\it fundamental
plane}, in the logarithmic space defined by the their effective
radius $R_{eff}$, intensity $I_e = I(R_{eff})$, and central
velocity dispersion $\sigma_0$ (Djorgovski \& Davis 1987, Dressler
et al. 1987, Bender et al. 1992, Burstein et al.
1997). This fundamental plane is usually parameterized as\,:

\begin{displaymath}
R_{eff} \propto \sigma_0^a I_e^b
\end{displaymath}
with $(a, b)$ predicted to be $(2, 1)$ if ETGs are in virial
equilibrium. The observed plane is, however, tilted with respect
to the virial one since the different determinations of $(a, b)$,
depending on the photometric band and the sample used, are always
different from the virial values. Jorgensen et al (1996) first
derived $a = 1.24 \pm 0.07$ and $b = -0.82 \pm 0.02$ from a set of
225 early\,-\,type galaxies in nearby clusters observed in the
r\,-\,band. While this result is consistent with the original
observations of Djorgovsky \& Davis (1987) and Dressler et al.
(1987), it is nevertheless in striking contrast with the most
recent determination relying on $\sim 9000$ ETGs observed within
the framework of the Sloan Digital Sky Survey (SDSS). Using this
large sample, Bernardi et al. (2003a) have found $a = 1.49 \pm
0.05$ and $b = -0.75 \pm 0.01$, which are more similar to the
K\,-\,band fundamental plane of Pahre et al. (1998). Although part
of this discrepancy may be alleviated taking care of the selection
effects and the fitting method adopted, the reason for such an
inconsistency is not currently clear.

While the precise values of the FP coefficients are still debated,
it is nevertheless clear that the observed FP is tilted with
respect to the virial prediction. Such a tilt could be caused by a
variation in the dynamical mass\,-\,to\,-\,light ratio for ETGs as
a result of a varying dark matter fraction (e.g., Padmanabhan et
al. 2004; Boylan\,-\,Kolchin et al. 2005) or stellar population
variations (e.g., Gerhard et al. 2001). It is also worth noting
that non\,-\,homology in the surface brightness profiles of
elliptical galaxies (e.g.,Graham \& Colless 1997; Trujillo et al
2004) may be an other explanation of the fundamental plane tilt.

Projections of the fundamental plane are also of interest in
studies of galaxy evolution. The SDSS team (Bernardi et al. 2003b)
measured both the radius\,-\,luminosity and the Faber\,-\,Jackson
(1976) relations obtaining respectively\,:

\begin{displaymath}
R_{eff} \propto L^{0.630 \pm 0.025} \ ,
\end{displaymath}

\begin{displaymath}
\sigma \propto L^{0.250 \pm 0.012} \ .
\end{displaymath}
Converting stellar luminosity in stellar masses according to the
prescription of Kauffmann et al. (2003), Shen et al. (2003) then
determined a size\,-\,mass relation as\,:

\begin{displaymath}
R_{eff} \propto M_{\star}^{0.56}
\end{displaymath}
for ETGs with $M_{\star} \ge 2 \times 10^{10} \ h^{-1} \ {\rm
M_{\odot}}$. Finally, the Kormendy relation (1985) between
effective surface brightness $\mu_e$ and effective radius has been
measured as\,:

\begin{displaymath}
\mu_e = (2.84  \pm 0.06) \log{R_{eff}} + (15.647 \pm 0.032) \ .
\end{displaymath}
It is worth noting that, although the FP itself is quite tight
over a wide range of ETGs (e.g., Bernardi et al 2003a), there are
a number of indications of variations in the FP projections
themselves with galaxy luminosity. Both the FP and its projections
(and their eventual correlation with luminosity) provide strong
constraints to any theory of formation and evolution of these
galaxies. Scaling relations are indeed often considered as major
constraints for models of galaxy formation and evolution. For
instance, they can be used to trace the contribution of the
different stellar components to the total luminosity of galaxies,
as in the case of the color\,-\,magnitude relation (see, e.g.,
Visvanathan \& Sandage 1977; Bower et al. 1992 for ETGs,  and
Tully et al. 1982, Gavazzi et al. 1996 for late\,-\,type galaxies)
or to study the relationship between kinematical, structural and
stellar population properties of galaxies as in the case of the
Tully\,-\,Fisher relation for spirals (Tully \& Fisher 1977) and
the fundamental plane for ellipticals. Moreover, the study of
these different scaling relations has been crucial for showing the
role of mass in the formation of galaxies (Gavazzi et al. 1996;
Boselli et al. 2001), a result now generally called downsizing
effect, which is a new major constraint for hierarchical models of
galaxy evolution (De Lucia et al. 2006).

A particularly interesting role has been played by the Kormendy
relation in elucidating important differences between giant and
dwarf ellipticals. Dwarf ellipticals (dEs) are considered to be
dark matter dominated systems and are therefore of vital interest
in understanding galaxy formation in general (see, e.g., Ferguson
\& Binggeli 1994 for a comprehensive review). Low luminosity
elliptical galaxies are distinguished from late\,-\,type galaxies
by their smooth surface brightness profiles. For systems with
luminosity smaller than $M_B \simeq -18$, the smooth profile
galaxies divide into two classes\,: compact galaxies with high
central surface brightness (e.g. M32), and diffuse galaxies with
low central surface brightness (e.g., the Local Group dwarf
spheroidals). In the hierachical galaxy formation scenario dEs are
supposed to be formed from average amplitude density fluctuations
(Dekel \& Silk 1986) where supernova driven galactic winds expel
the gas content and reshape the galaxy body. In an alternative
view, dEs could form from progenitors through the process of
galaxy harassment (Moore et al. 1998). Likely candidates for the
progenitors could be stripped late type galaxies. Data from the
Local Group dEs and few galaxies in the Virgo cluster (Bender \&
Nieto, 1990; Held et al. 1990) indicated
that dEs are supported by anisotropy thus introducing a dichotomy
in the otherwise linear sequence of increasing rotational support
with decreasing luminosity for the whole class of ellipticals.

Furthermore, the study of the B band surface brightness vs.
absolute magnitude relation and of the Kormendy relation has
originally shown a strong, apparent dichotomy in the behavior of
dwarf and giant ellipticals. While in dwarfs the effective or
central surface brightness increases with luminosity, the opposite
trend is seen in giants (e.g. Ferguson \& Binggeli 1994; Graham \&
Guzman 2003). An opposite trend between giants and dwarfs has been
also observed in the Kormendy relation (e.g. Kormendy 1985;
Capaccioli et al. 1992). This surprising result has been
originally interpreted as a clear indication that dEs are not the
low luminosity extension of giants, but rather an independent
class of objects. The study of Graham \& Guzman (2003), based on
HST data, has however shown that this dichotomy is only apparent
since it is due to a gradual steepening of the central radial
profile with luminosity. On the other hand, several recent
observational evidences and simulations seem to indicate that
local group dwarf spheroidals (Mayer et al. 2006), Virgo cluster
dEs (Barazza et al. 2002; Conselice et al. 2003; van Zee et al.
2004; Mastropietro et al. 2005; Lisker et al. 2006a,b; 2007, 2008,
Lisker \& Han 2008; Michielsen et al. 2008) or generally dwarf
spheroidals in other clusters such as Coma (Smith et al. 2008),
Perseus (Penny \& Conselice 2008) or in the SDSS (Haines et al.
2007) might be late\,-\,type galaxies recently perturbed by an
hostile environment through ram pressure stripping or galaxy
harassment.

It is of peculiar interest the scaling relation connecting
effective radius and mass in dEs and E galaxies. As noticed by
Shen et al. (2003), this scaling relation in the case of E is
$R_{eff} \propto M^{0.56}$, while in dwarfs the slope decreases to
$R_{eff} \propto M^{0.14}$. The same behavior has been observed
by Dabringhausen et al. (2008) thus arguing in favour of a
different origin, such as the dE galaxies being tidal-dwarf
galaxies (Okazaki \& Taniguchi 2000). If this were true, then dE
galaxies and E galaxies, that presumably formed from a monolithic
and rapid collapse, would have fundamentally different phase space
distribution functions (DFs). One way to study a possible
indication for a dynamically different origin of dE and E galaxies
is to solve for a possible indication of different angular
momentum dependency of the DF. With this contribution we perform a
first step in this direction by assuming spherically symmetric
models for dE and E galaxies, but allowing the DF to depend on the
angular momentum. Angular momentum is acquired through tidal
torques of the protogalaxy with the neighboring ones (Hoyle 1949;
Peebles 1969; Hoffman 1986a,b; Eisenstein \& Loeb
1995; Catelan \& Theuns 1996a,b). Baryons and the halo particles
are well mixed initially (Mo et al. 1998; Cardone \& Sereno 2005;
Treu \& Koopmans 2002; Keeton 2001). This is due to the fact that
the original angular momentum of the dark matter halo comes from
gravitational (tidal) interactions with its environment. Thus, the
dark matter and the gas experience the same torque in the process
of halo assembly and should initially have (almost) the same
specific angular momentum (Klypin et al. 2002).

In the following, we shall study how an injection of angular
momentum in the system changes the scaling relation
$R_{eff}$\,-\,$M_{\star}$. In a spherically symmetric isotropic
system, one expects that $R_eff \propto  M_{\star}^{\gamma}$, with
$\gamma = 1/2$, while observations give $\gamma = 0.56$ (Shen et
al. 2003). We therefore investigate whether the injection of
angular momentum may help in explaining the observed deviation
from the naive expectation.

The paper is organized as follows. Section 2 describes the Sersic
profile and the Prugniel and Simien model used to deproject it.
Sections 3 and 4 are devoted to the calculation of the model DF in
the isotropic and anisotropic case, respectively. The resulting
size\,-\,mass relation is discussed in Sect. 5, while we derive a
similar relation for dark halo models in Sect. 6. Conclusions are
then presented in Sect. 7.

\section{The PS model}

Notwithstanding the large range in luminosity, mass and size, ETGs
represent a surprisingly regular class of objects concerning their
photometric properties. Indeed, as well known
\cite{CCD93,GC97,PS97}, their surface brightness is well described
by the Sersic (1968) profile\,:

\begin{equation}
I(R) = I_e \exp{\left \{ - b_n \left [ \left ( \frac{R}{R_{eff}}
\right )^{1/n} - 1 \right ] \right \}} \label{eq: ir}
\end{equation}
with $R$ the cylindrical radius\footnote{Note that we have
implicitly assumed that the intensity $I$ does not depend on the
angular coordinates. Actually, the isophotes are not concentric
circles, but rather ellipses with variable ellipticities and
position angles so that $I = I(R, \varphi)$. However, in order to
be consistent with our assumption of spherical symmetry of the
three dimensional mass profile, we will neglect such an effect
and, following a common practice, {\it circularize} the intensity
profile considering circular isophothes with radii equal to the
geometric mean of the major and minor axes.} on the plane of the
sky and $I_e$ the luminosity intensity at the effective radius
$R_{eff}$. The constant $b_n$ is determined by the condition that
the luminosity within $R_{eff}$ is half the total luminosity, i.e.
$b_n$ is found by solving\,:

\begin{equation}
\Gamma(2n, b_n) = \Gamma(2n)/2 \label{eq: bn}
\end{equation}
where $\Gamma(a, z)$ is the incomplete $\Gamma$ function. Although
Eq.(\ref{eq: bn}) is straightforward to solve numerically, a very
good analytical approximation is given by \cite{CB99}\,:

\begin{displaymath}
b_n = 2n - \frac{1}{3} - \frac{0.009876}{n} \ .
\end{displaymath}
The deprojection of the intensity profile in Eq.(\ref{eq: ir}) is
straightforward under the hypothesis of spherical symmetry, but,
unfortunately, the result turns out to be a somewhat involved
combinations of the unusual Meijer functions \cite{MC02}. In order
to not deal with these difficult to handle expression, we prefer
to use the model proposed by Prugniel and Simien (1997, hereafter
PS97) whose three dimensional luminosity density reads\,:

\begin{equation}
j(r) = j_0 \left ( \frac{r}{R_{eff}} \right )^{-p_n} \exp{\left [
-b_n \left ( \frac{r}{R_{eff}} \right )^{1/n} \right ]} \label{eq:
jr}
\end{equation}
with

\begin{equation}
j_0 = \frac{I_0 b_n^{n (1 - p_n)}}{2 R_{eff}}
\frac{\Gamma(2n)}{\Gamma[n (3 - p_n)]} \ . \label{eq: jz}
\end{equation}
Here, $I_0 = I(R = 0) = I_e {\rm e}^{b_n}$, while the constant
$p_n$ is chosen so that the projection of Eq.(\ref{eq: jr})
matches a Sersic profile with the same values of $(n, R_{eff},
I_e)$. A useful fitting formula is given as \cite{Metal01}\,:

\begin{displaymath}
p_n = 1.0 - \frac{0.6097}{n} + \frac{0.00563}{n^2} \ .
\end{displaymath}
Because of the assumed spherical symmetry, the luminosity profile
may be simply obtained as\,:

\begin{displaymath}
L(r) = 4 \pi \int_{0}^{r}{r'^2 j(r') dr'}
\end{displaymath}
which, for the PS model, becomes\,:

\begin{equation}
L(r) = L_T {\times} \frac{\gamma[n (3 - p_n), b_n
\eta^{1/n})}{\Gamma[n (3 - p_n)]} \label{eq: lr}
\end{equation}
where the total luminosity $L_T$ reads\,:

\begin{equation}
L_T = 2 \pi n b_n^{-2n} {\rm e}^{b_n} \Gamma(2n) I_e R_{eff}^2 \ .
\label{eq: lt}
\end{equation}
Note that the total luminosity is the same as the projected one
for the corresponding Sersic profile as can be immediately check
computing\,:

\begin{displaymath}
L_T = 2 \pi \int_{0}^{\infty}{I(R) R dR} \ .
\end{displaymath}
As a final remark, let us stress that, under the hypothesis of
constant $M/L$ ratio, we can convert the luminosity density $j(r)$
in a mass density $\rho(r)$ simply as $\rho(r) = \Upsilon_{\star}
j(r)$ so that the total mass of the stellar component reads
$M_{\star} = \Upsilon_{\star} L_T$. This may be determined from
the measurement of the photometric parameters $(n, R_{eff}, I_e)$
provided that an estimate of the stellar $M/L$ ratio
$\Upsilon_{\star}$ is available (for instance, from the relation
between $\Upsilon_{\star}$ and the colors or from fitting the
galaxy spectrum to stellar population synthesis models). It is,
finally, worth stressing that, according to Eq.(\ref{eq: lt}), we
expect a size\,-\,mass relation as $R_{eff} \propto
M_{\star}^{1/2}$ as a consequence of the total luminosity being
the same as the projected one. Actually, some deviations from this
simple scaling may be come out if we give off the homology hypothesis, i.e.
that the Sersic index $n$ is the same for all the galaxies. Indeed, should 
$n$ systematically varies with the total stellar mass $M_{\star}$, e.g. as 
$n \propto M_{\star}^{\nu}$, then Eq.(\ref{eq: lt}) gives $R_{eff} \propto
M_{\star}^{(1 - \nu)/2}$. However, the scaling $n \propto M_{\star}^{\nu}$ is 
far to be verified so that we will assume that the homology hypothesis indeed
holds, at least as a first approximation.

\section{The isotropic DF}

Assuming isotropy in the velocity space, the distribution function
(hereafter, DF) of a spherically symmetric model may be easily
recovered resorting to the Eddington formula \cite{BT87}\,:

\begin{eqnarray}
f({\cal{E}}) & = & \frac{1}{\sqrt{8} \pi^2} \frac{d}{d{\cal{E}}}
\int_{0}^{{\cal{E}}}{\frac{d\rho}{d\Psi} \frac{d\Psi}{\sqrt{{\cal{E}} - \Psi}}} \nonumber \\
~ & = & \frac{1}{\sqrt{8} \pi^2}
\int_{0}^{{\cal{E}}}{\frac{d^2\rho}{d\Psi^2}
\frac{d\Psi}{\sqrt{{\cal{E}} - \Psi}}} +
\frac{1}{\sqrt{{\cal{E}}}} \left ( \frac{d\rho}{d\Psi} \right
)_{\Psi = 0} \label{eq: eddfor}
\end{eqnarray}
with ${\cal{E}} = \Psi - v^2/2$ the total energy per unit of mass
and $\Psi(r) = - \Phi(r)$ the gravitational potential (with the
sign changed). Note that we will refer to the second expression
since it avoids differentiating the integral which has often to be
evaluated numerically. Moreover, the second term in this latter
expression is typically negligible because the mass density scales
faster than $r^{-2}$ in most models.

As a preliminary step, we have therefore to evaluate $\Phi(r)$ by
solving the Poisson equation $\nabla^2 \Phi = 4 \pi G \rho$. Due
to the spherical assumption, this reduces to\,:

\begin{displaymath}
\frac{1}{r^2} \frac{d}{dr} \left( r^2 \frac{d\Phi}{dr} \right ) =
4 \pi G \rho \ .
\end{displaymath}
It is convenient to introduce the following dimensionless
quantities\,:

\begin{displaymath}
\eta = r/R_{eff} \ \ , \ \ \tilde{\Phi} = \Phi/4 \pi G \rho_0
R_{eff}^2
\end{displaymath}
with $\rho_0 = \Upsilon_{\star} j_0$ so that the Poisson equation
becomes\,:

\begin{equation}
\frac{1}{\eta^2} \frac{d}{d\eta} \left ( \eta^2
\frac{d\tilde{\Phi}}{d\eta} \right ) = \eta^{-p_n} \exp{[-b_n
(\eta^{1/n} - 1)]} \ . \label{eq: poisson}
\end{equation}
Integrating first and imposing that the force $F \propto
-d\tilde{\Phi}/d\eta$ vanishes at the centre gives\,:

\begin{equation}
\frac{d\tilde{\Phi}}{d\eta} = n b_n^{-n(3 - p_n)} \gamma[n(3 -
p_n), b_n \eta^{1/n}]/\eta^2 \label{eq: dphideta}
\end{equation}
so that, after a second integration, we finally get\,:

\begin{eqnarray}
\tilde{\Phi}(\eta) & = & - n b_n^{-n(3 - p_n)} \gamma[n(3 - p_n), b_n \eta^{1/n}]/\eta \nonumber \\
~ & - & n b_n^{-n(2 - p_n)} \Gamma[n(2 - p_n), b_n \eta^{1/n}]
\label{eq: phi}
\end{eqnarray}
having imposed that the potential is null at infinity. We stress
that the central scaled potential only depends on $n$\,:

\begin{equation}
\lim_{\eta \rightarrow 0}{\tilde{\Phi}(\eta)} = -n b_n^{-n(2 -
p_n)} \Gamma[n(2 - p_n)] \ . \label{eq: psizero}
\end{equation}
In order to use the Eddington formula, we should now invert the
relation $\Psi = \Psi(r)$ and replace $r(\Psi)$ into Eq.(\ref{eq:
jr}) to get $\rho(\Psi)$ and then $d\rho/d\Psi$. Needless to say,
this is not analytically possible so that we prefer to escape the
problem by a simple change of variables. First, let us consider\,:

\begin{eqnarray}
\frac{1}{\sqrt{{\cal{E}}}} \left ( \frac{d\rho}{d\Psi} \right
)_{\Psi = 0} & = &
\frac{1}{\sqrt{4 \pi G \rho_0 R_{eff}^2 \tilde{{\cal{E}}}}} \nonumber \\
~ & {\times} & \left [ 4 \pi G R_{eff}^2 \left (
\frac{d\tilde{\Psi}}{d\eta} \right )^{-1} \left (
\frac{d\tilde{\rho}}{d\eta} \right ) \right ] \nonumber \\ ~ & = &
\left ( \frac{1}{4 \pi G \rho_0^{1/3} R_{eff}^2} \right )^{3/2}
\nonumber \\ ~ & {\times} & \frac{1}{\sqrt{\tilde{{\cal{E}}}}}
\lim_{\eta \rightarrow \infty}{\left [ \left (
\frac{d\tilde{\Psi}}{d\eta} \right )^{-1} \left (
\frac{d\tilde{\rho}}{d\eta} \right ) \right ]} \label{eq: st}
\end{eqnarray}
having trivially defined $\tilde{\Psi} = \Psi/4 \pi G \rho_0
R_{eff}^2 = -\tilde{\Phi}$, $\tilde{{\cal{E}}} = {\cal{E}}/4 \pi G
\rho_0 R_{eff}^2$, $\tilde{\rho} = \rho/\rho_0$, and used the
property that $\Psi$ vanishes at infinity. We can now proceed the
same way to get\,:

\begin{eqnarray}
\frac{d^2 \rho}{d\Psi^2} & = & \left ( \frac{1}{4 \pi G
\rho_0^{1/2} R_{eff}^2} \right )^2 \left (
\frac{d\tilde{\Psi}}{d\eta} \right )^{-2} \nonumber \\ ~ &
{\times} & \left \{ \frac{d^2\tilde{\rho}}{d\eta^2} - \left (
\frac{d\tilde{\Psi}}{d\eta} \right )^{-1} \left (
\frac{d^2\tilde{\Psi}}{d\eta^2} \right )
\frac{d\tilde{\rho}}{d\eta} \right \} \ . \label{eq: ft}
\end{eqnarray}
Inserting Eqs.(\ref{eq: st}) and (\ref{eq: ft}) into the second of
Eq.(\ref{eq: eddfor}), the isotropic DF finally reads\,:

\begin{equation}
f({\cal{E}}) = \frac{\tilde{f}(\tilde{{\cal{E}}})}{G^{3/2}
\rho_0^{1/2} R_{eff}^3} \label{eq: eddbis}
\end{equation}
with the dimensionless DF defined as\,:

\begin{eqnarray}
\tilde{f}(\tilde{{\cal{E}}}) & = & \frac{1}{2^{9/2} \pi^{7/2}}
\nonumber \\ ~ & {\times} & \int_{\tilde{\eta}}^{\infty} {\left (
\frac{d\tilde{\Psi}}{d\eta} \right )^{-1} \left \{ \left (
\frac{d\tilde{\Psi}}{d\eta} \right )^{-1} \left (
\frac{d^2\tilde{\Psi}}{d\eta^2} \right )
\frac{d\tilde{\rho}}{d\eta} - \frac{d^2\tilde{\rho}}{d\eta^2}
\right \}} \nonumber \\ ~ & + & \left ( \frac{1}{4 \pi} \right
)^{3/2} \frac{1}{\sqrt{\tilde{{\cal{E}}}}} \lim_{\eta \rightarrow
\infty}{\left [ \left ( \frac{d\tilde{\Psi}}{d\eta} \right )^{-1}
\left ( \frac{d\tilde{\rho}}{d\eta} \right ) \right ]} \label{eq:
tildedf}
\end{eqnarray}
with $\tilde{\eta}$ the (numerical) solution of\,:

\begin{figure}
\includegraphics[width=8.5cm]{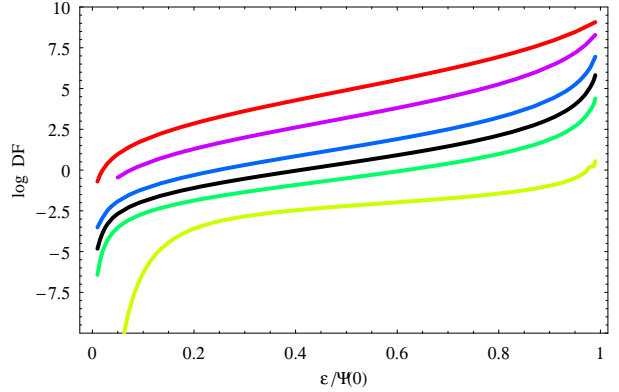}
\caption{The normalized isotropic DF $\tilde{f}(\tilde{{\cal{E}}},
n)$ as function of the normalized energy
$\tilde{{\cal{E}}}/\tilde{\Psi}(0)$ for $n$ ranging from 1 to 9 in
steps of 2 (from the bottom yellow to the top red curve) with the
black line referring to the de Vaucouleurs ($n = 4$) case. Note
that, due to numerical reason, $\tilde{\Psi}(0)$ is actually
$\tilde{\Psi}(10^{-4})$ without any significant loss of
precision.} \label{fig: dfiso}
\end{figure}

\begin{equation}
\tilde{\Psi}(\tilde{\eta}) = \tilde{{\cal{E}}} \ . \label{eq:
etatilde}
\end{equation}
Using Eqs.(\ref{eq: jr}) and (\ref{eq: phi}), it is only a matter
of algebra to compute the different terms entering Eq.(\ref{eq:
tildedf}) for the PS model. We do not report here the full
expressions for sake of shortness, but we stress that the second
term identically vanishes, while the final DF is a function of the
scaled energy $\tilde{{\cal{E}}}$ and the slope $n$ of the Sersic
profile. Moreover, since putting together Eqs.(\ref{eq: jz}) and
(\ref{eq: lt}) gives\,:

\begin{equation}
\rho_0 = \Upsilon_{\star} j_0 = \frac{M_{\star}}{4 \pi R_{eff}^3}
\frac{b_n^{n(3 - p_n)}}{n \Gamma[n(3 - p_n)]} \ , \label{eq:
rhozero}
\end{equation}
we can rewrite the isotropic DF for the PS model as\,:

\begin{equation}
f({\cal{E}}) = \frac{\tilde{f}(\tilde{{\cal{E}}}, n)}{G^{3/2}
\Upsilon_{\star}^{1/2} j_0^{1/2} R_{eff}^3} = 
\frac{\lambda(n) \tilde{f}(\tilde{{\cal{E}}}, n)}{G^{3/2}
M_{\star}^{1/2} R_{eff}^{3/2}} 
\label{eq: isodf}
\end{equation}
with $I_0 = I(R = 0) = I_e {\rm e}^{b_n}$ and\,:

\begin{equation}
\lambda(n) = \left \{ \frac{n \Gamma[4 \pi n(3 - p_n)]}{b_n^{n(3 -
p_n)}} \right \}^{1/2} \ .
\label{eq: deflambda}
\end{equation}
Fig.\,\ref{fig: dfiso} shows the normalized DF
$\tilde{f}(\tilde{{\cal{E}}}, n)$ as function of
$\tilde{{\cal{E}}}$ over the energy range
$\tilde{\Psi}(\eta_{min}) \le \tilde{{\cal{E}}} \le
\tilde{\Psi}(\eta_{max})$ with $(\eta_{min}, \eta_{max}) =
(10^{-4}, 10^{4})$. It is worth noting that the resulting DF is
always positive that is to say the PS model is a physical one
whatever the value of $n$ is.

\section{The anisotropic DF}

According to the model we are investigating, the PS density profile
and DF describe the initial configuration of the baryons which will then 
originate the visible part of the galaxy. Let us now assume that a physical 
mechanism takes place injecting angular momentum into this initial configuration. 
Such an injection can be attributed, for instance, to the action of tidal torques 
between neighboring protogalaxies. As the protogalaxy collapses, it spins more and 
more rapidly thus acquiring angular momentum. As a further possibility, it is worth 
remembering that a merger of two disc galaxies also produces a spheroidal stellar system
retaining angular momentum from the original rotating discs. Investigating the details
of how angular momentum is injected in the initial PS\,-\,like configuration is outside
our aims here, but we just recall these two possible scenarios to show that such
a transfer of angular momentum is indeed possible. As a consequence, the phase space 
distribution of stars will be altered thus deviating from the isotropic DF 
we have considered above.

For spherically anisotropic systems, the DF may depend not only on
the energy ${\cal{E}}$, but also on the total angular momentum
$L$. In particular, we will focus our attention on the
Osipkov\,-\,Merritt (OM) models \cite{O79,M85} where the DF
reads\,:

\begin{equation}
f({\cal{E}}, L) = f_0(Q) L^{2 \alpha} \label{eq: omdf}
\end{equation}
where the isotropic DF $f_0$ is evaluated in the lowered binding
energy $Q = {\cal{E}} - L^2/2 r_a^2$. The slope parameter $\alpha$
and the anisotropy radius $r_a$ determine the velocity dispersion
anisotropy profile given by\,:

\begin{equation}
\beta(r) = \frac{r^2 - \alpha r_a^2}{r^2 + r_a^2} \ . \label{eq:
beta}
\end{equation}
For a positive $\alpha$, the anisotropy is tangential for $r^2 <
\alpha r_a^2$ to become then radial for $r^2/r_a^2 > \alpha$. On
the other hand, a negative $\alpha$ gives a model which is
radially anisotropic everywhere whatever the value of $r_a$ is.
Note that, for $\alpha = 0$ and $r_a = 1$, we obtain a model which
is isotropic in the inner region and radially anisotropic in the
outer ones.

The density profile corresponding to the anisotropic OM models may
be computed as \cite{C91}\,:

\begin{eqnarray}
\rho(r) & = & \frac{(2 \pi)^{3/2} (2 r^2)^{\alpha}}{(1 +
r^2/r_a^2)^{\alpha + 1}} \frac{\Gamma(\alpha + 1)}{\Gamma(\alpha +
3/2)} \nonumber \\ ~ & {\times} & \int_{0}^{\Psi} {f_0(Q) (\Psi -
Q)^{\alpha + 1/2} dQ} \ . \label{eq: cudd}
\end{eqnarray}
Inserting Eq.(\ref{eq: isodf}) for the isotropic DF and changing
variable from $Q$ to $\tilde{Q} = Q/4 \pi G \rho_0 R_{eff}^2$, we
easily get\,:

\begin{eqnarray}
\rho(\tilde{\Psi}, \eta) & = & \frac{(2 \pi)^{3/2} (2
R_{eff}^2)^{\alpha} \eta^{2 \alpha}} {(1 +
\eta^2/\eta_a^2)^{\alpha + 1}} \frac{\Gamma(\alpha +
1)}{\Gamma(\alpha + 3/2)} \nonumber \\  ~ & {\times} & \frac{(4
\pi G \rho_0 R_{eff}^2)^{\alpha + 3/2}}{G^{3/2} \rho_0^{1/2}
R_{eff}^3} \int_{0}^{\tilde{\Psi}}{\tilde{f}(\tilde{Q}, n)
(\tilde{\Psi} - \tilde{Q})^{\alpha + 1/2} d\tilde{Q}} \nonumber \\
~ & = & \frac{2^{(6\alpha + 9)/2} \pi^{\alpha + 3} \Gamma(\alpha +
1)}{\Gamma(\alpha + 3/2) \tilde{\Psi}_0^{-(\alpha + 3/2)}} \
{\times} \ G^\alpha \rho_0^{\alpha + 1} R_{eff}^{4 \alpha}
\nonumber \\ ~ & {\times} & \frac{\eta^{2\alpha}}{(1 +
\eta^2/\eta_a^2)^{\alpha + 1}} \int_{0}^{\psi}{\tilde{f}(q,
n)(\psi - q)^{\alpha + 1/2} dq} \label{eq: rhocudd}
\end{eqnarray}
where we have also introduced the more manageable variables $\psi
= \tilde{\Psi}/\tilde{\Psi}_0$, $q = \tilde{Q}/\tilde{\Psi}_0$,
with $\tilde{\Psi}_0 = \tilde{\Psi}(\eta = 0)$.

As an important step, we have to check that the right hand side of
Eq.(\ref{eq: rhocudd}) has indeed the physical dimensions of a
mass density. It is straightforward to verify that this is not the
case because of the term $(G \rho_0 R_{eff})^{\alpha}$. The origin
of this discrepancy may be, however, easily understood noting that
$G \rho_0 R_{eff}^4$ has the same physical dimensions as $L^2$,
i.e. the square modulus of the angular momentum. Actually, in
Eq.(\ref{eq: omdf}), we have implicitly assumed that $L$ is
dimensionless in order to follow the common practice. However,
more correctly, we must introduce an (up to now) arbitrary
normalization constant angular momentum $L_0$ to recover the
correct result. Taking care of this and using Eq.(\ref{eq:
psizero}) for $\tilde{\Psi}_0$, we therefore rewrite the mass
density as\,:

\begin{eqnarray}
\rho(\psi, \eta) & = & \frac{2^{(6\alpha + 9)/2} \pi^{\alpha + 3}
\Gamma(\alpha + 1)} {\Gamma(\alpha + 3/2)} \left \{ \frac{n
\Gamma[n(2 - p_n)]}{b_n^{n(2 - p_n)}} \right \}^{\alpha + 3/2}
\nonumber \\ ~ & {\times} &  \left ( \frac{L_{eff}}{L_0} \right
)^{2\alpha} \ \frac{\rho_0 \eta^{2 \alpha} \hat{\rho}(\psi, n,
\alpha)}{(1 + \eta^2/\eta_a^2)^{\alpha + 1}} 
\label{eq: rhocuddend}
\end{eqnarray}
having naively defined\footnote{It is worth stressing that
$L_{eff}$ is not the value of the angular momentum at the
effective radius. Although not completely correct, an order of
magnitude estimate of $L_{eff}$ is given by $\sqrt{G M_{\star}
R_{eff}}$, with $M_{\star}$ the total mass of the luminous
component.}\,:

\begin{equation}
L_{eff} = \sqrt{G \rho_0 R_{eff}^4} \ , \label{eq: defleff}
\end{equation}

\begin{equation}
\hat{\rho}(\psi, n, \alpha) = \int_{0}^{\psi}{\tilde{f}(q, n)
(\psi - q)^{\alpha + 1/2} dq} \ . \label{eq: defhatrho}
\end{equation}
Eq.(\ref{eq: rhocuddend}) gives $\rho$ as function of $\psi$ and
$\eta$, while one is usually interested in $\rho$ as function of
$\eta$ only. To this aim, we have to first solve the Poisson
equation for the scaled dimensionless potential $\psi$. Starting
from

\begin{displaymath}
\nabla^2 \Psi = -4 \pi G \rho \ ,
\end{displaymath}
it is immediate to get\,:

\begin{displaymath}
\nabla^2 \psi = \frac{1}{\eta^2} \frac{d}{d\eta} \left ( \eta^2
\frac{d\psi}{d\eta} \right ) = - \frac{\rho}{\tilde{\Psi}_0
\rho_0} \ ,
\end{displaymath}
so that we have to solve the following differential equation\,:

\begin{figure*}
\centering
\subfigure{\includegraphics[width=8cm]{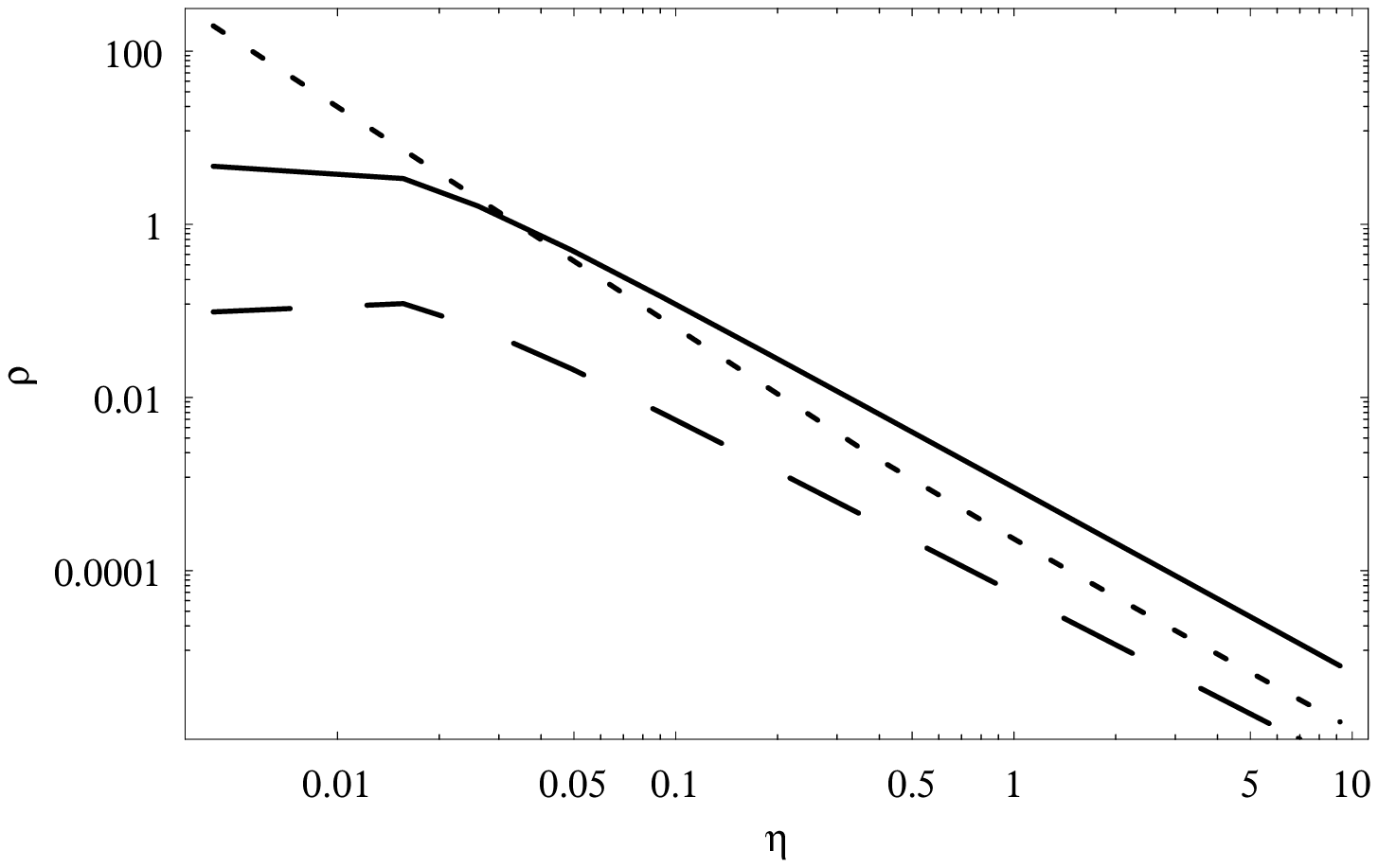}}
\goodgap \subfigure{\includegraphics[width=8cm]{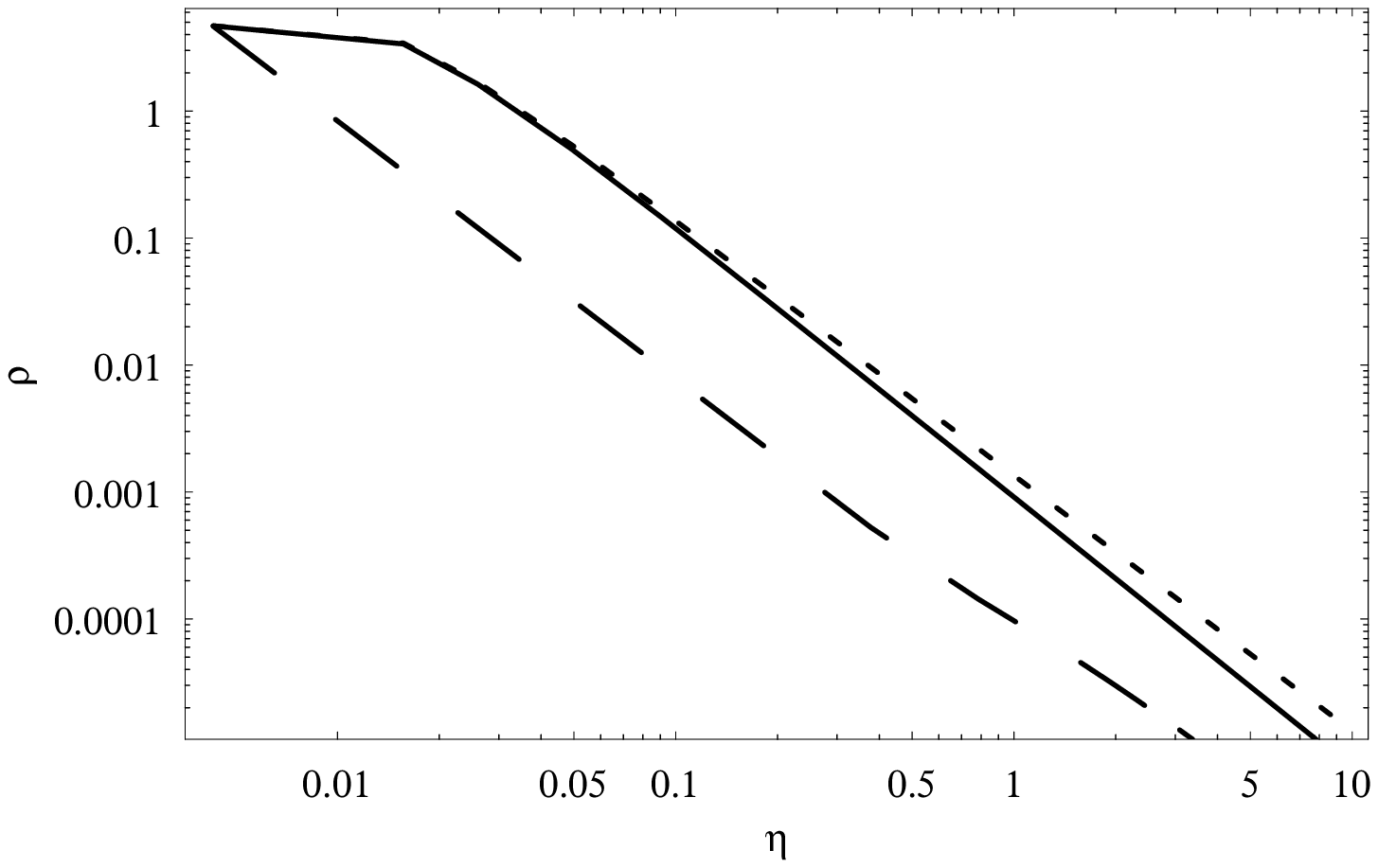}} \\
\caption{The shape of the density profile of the OM\,-\,PS model. In the left panel, we set 
$\log{(L_{eff}/L_0)} = 0.0$ and $\alpha = 0.15$ (short dashed), $0.50$ (solid), $0.75$ 
(long dashed), while in the right panel $\alpha = 0.50$ and three values of $\log{(L_{eff}/L_0)}$
are considered, namely $-1.0$ (short dashed), $0.0$ (solid), $1.0$ (long dashed). We 
arbitrarily set $n = 4$ and $\eta_a = r_a/R_{eff} = 0.01$ in both panels.}

\label{fig: rhocudd}
\end{figure*}

\begin{eqnarray}
\nabla^2 \psi & = & - \frac{2^{(6\alpha + 9)/2} \pi^{\alpha + 3}
\Gamma(\alpha + 1)} {\Gamma(\alpha + 3/2)} \left \{ \frac{n
\Gamma[n(2 - p_n)]}{b_n^{n(2 - p_n)}} \right \}^{\alpha + 1/2}
\nonumber \\ ~ & {\times} &  \left ( \frac{L_{eff}}{L_0} \right
)^{2\alpha} \ \frac{\eta^{2 \alpha} \hat{\rho}(\psi, n,
\alpha)}{(1 + \eta^2/\eta_a^2)^{\alpha + 1}} \ . \label{eq:
ompoisson}
\end{eqnarray}
Needless to say, solving this equation analytically is not
possible given its high nonlinearity because of the way $\psi$
enters in the determination of $\hat{\rho}(\psi, n, \alpha)$.
However, a numerical solution is possible provided one sets the
Sersic index $n$, the anisotropy parameters $(\alpha, \eta_a)$,
and the scaled characteristic angular momentum\footnote{Note that
the actual value of $L_0$ is up to now meaningless since what is
needed to solve Eq.(\ref{eq: ompoisson}) is the ratio
$L_{eff}/L_0$. Changing the value of $L_0$ gives only a trivial
rescaling of the results.} $L_{eff}/L_0$. Once such a solution for
$\psi$ has been obtained, we can numerically invert the relation
$\psi = \psi(\eta)$ to finally get $\rho = \rho(\eta)$. It is,
however, worth stressing that, as a consequence of how
$\psi(\eta)$ has been obtained, the mass density $\rho(\eta)$ will
be parameterized by the quantities $(n, \alpha, \eta_a,
L_{eff}/L_0)$ which are therefore the four parameters needed to
assign the anisotropic OM\,-\,PS model.

It is interesting to look at the density profile of the OM\,-\,PS model. 
To this end, we show in Fig.\,\ref{fig: rhocudd} the normalized density
$\rho_s = \eta^{2 \alpha}/(1 + \eta^2/\eta_a^2)^{\alpha + 1} \hat{\rho}(\eta)$
for models with $n = 4$ arbitrarily setting $\eta_a = 0.01$. As a general result,
we note that an inner core developes when adding angular momentum to 
the orbit distribution. As left panel shows, the higher is $\alpha$, the more
prominent is the core, i.e. $\rho_s \sim const$ for $\eta < \eta_c$ with $\eta_c$
an increasing function of $\alpha$. However, outside this inner region, the model
resembles the original PS one so that it is likely that its projection still recovers
the Sersic surface brightness profile with great precision everywhere but for $\eta
< \eta_c$. Since $\eta_c$ is typically quite small $(\eta_c \sim 0.01)$, we can
still use the value of $n$ retrieved from fitting the Sersic law to the observed 
photometry, which is what we will assume hereafter. Concerning the role of $L_{eff}/L_0$,
right panel shows that it is mainly a scaling parameter shifting up the full profile as 
$L_{eff}/L_0$ becomes smaller\footnote{Note that, in Fig.\,\ref{fig: rhocudd}, we have 
used $\log{(L_{eff}/L_0)}$ as variable so that we are actually investigating a range 
covering two order of magnitudes. For this reason, the inner core of the model
with $\log{(L_{eff}/L_0)} = 1.0$ shifts outside the plot, but it is still present.}.
While changing $n$ leaves qualitatively unaltered the above results, the value 
adopted for $\eta_a$ plays a most significant role. Indeed, depending on the $(\alpha,
L_{eff}/L_0)$, should $\eta_a$ be larger than a critical value, $\rho_s(\eta)$ can 
also become an increasing function of $\eta$ in the very inner regions thus leading to 
an unphysical density profile. Investigating in detail what is the region of 
the 4D parameter space $(n, \alpha, \eta_a, L_{eff}/L_0)$ giving rise to physically 
meaningful OM\,-\,PS models is, however, outside our aims here. As a final remark, we 
warn the reader to not use Eq.(\ref{eq: rhozero}) to replace $\rho_0$ with the total 
mass $M_{\star}$ in Eq.(\ref{eq: defleff}) since Eq.(\ref{eq: rhozero}) only holds
when $\rho(r)$ has the functional expression given by Eq.(\ref{eq:
jr}) which is not the case here. Nevertheless, one can still resort to 
Eq.(\ref{eq: jz}) since this is an outcome of the requirement that the projected 
PS model matches the Sersic profile which is still approximately true for the OM\,-\,PS case 
(except in the very inner regions).

\section{The size\,-\,mass relation}

Notwithstanding the anisotropy in the velocity space, the
spherical symmetry of the system makes it possible to compute the
total mass as\,:

\begin{displaymath}
M_{\star} = 4 \pi \int_{0}^{\infty}{r^2 \rho(r) dr} = 4 \pi
R_{eff}^3 \int_{0}^{\infty}{\eta^2 \rho(\eta) d\eta} \ .
\end{displaymath}
It is now only a matter of algebra to insert Eq.(\ref{eq:
rhocudd}) into the above relation to get\,:

\begin{eqnarray}
M_{\star} & = & \frac{2^{(6\alpha + 13)/2} \pi^{\alpha + 4}
\Gamma(\alpha + 1)}{\Gamma(\alpha + 3/2)} \left \{ \frac{{\rm
e}^{b_n} b_n^{n(1 - p_n)} \Gamma(2n)}{\Gamma[n(3 - p_n)]} \right
\} \nonumber \\ ~ & {\times} & \left \{ \frac{n \Gamma[n(2 -
p_n)]}{b_n^{n(2 - p_n)}} \right \}^{\alpha + 3/2} \left (
\frac{L_{eff}}{L_0} \right )^{2 \alpha} \Upsilon_{\star} I_e
R_{eff}^2 \nonumber \\ ~ & {\times} & {\cal{I}}(n, \alpha, \eta_a,
L_{eff}/L_0) \ , \label{eq: massstar}
\end{eqnarray}
where we have used $\rho_0 = \Upsilon_{\star} j_0$ with $j_0$ from
Eq.(\ref{eq: jz}) and defined\,:

\begin{equation}
{\cal{I}}(n, \alpha, \eta_a, L_{eff}/L_0) =
\int_0^{\infty}{\frac{\eta^{2\alpha + 2} \hat{\rho}(\eta)}{(1 +
\eta^2/\eta_a^2)^{\alpha + 1}} d\eta} \ . \label{eq: defmassint}
\end{equation}
Note that, when $\alpha = 0$, the density profile reduces to the
original PS one and we get the usual result $M_{\star} \propto
R_{eff}^2$, i.e. $\log{R_{eff}} \propto (1/2) \log{M_{\star}}$, as
expected.

Let us now suppose that, for a fixed value of $n$, the mass
integral (\ref{eq: defmassint}) can be approximated as\,:

\begin{eqnarray}
\log{{\cal{I}}} & = & a + b \alpha +
c \log{{\cal{L}}_{eff}} + d \log{\eta_a} \nonumber \\
~ & + & e \alpha \log{{\cal{L}}_{eff}} + f \alpha \log{\eta_a}
\label{eq: ansatz}
\end{eqnarray}
with ${\cal{L}}_{eff} = L_{eff}/L_0$ and $(a, b, c, d, e, f)$
constant parameters to be determined. Inserting this ans\"atz into
Eq.(\ref{eq: massstar}) gives\,:

\begin{eqnarray}
M_{\star} & = & \frac{2^{(6 \alpha + 13)/2} \pi^{\alpha + 4}
\Gamma(\alpha + 1)}{\Gamma(\alpha + 3/2)} \left \{
\frac{\rm{e}^{b_n} b_n^{n(1 - p_n)} \Gamma(2n)}{\Gamma[n(3 -
p_n)]}
\right \} \nonumber \\
~ & \times & \left \{ \frac{n \Gamma(2n)}{b_n^{n(2 - p_n)}} \right \}^{\alpha + 3/2} \Upsilon_{\star} I_e R_{eff}^2 \nonumber \\
~ & \times & {\cal{A}}(\alpha) \ \eta_a^{d + f \alpha} \
{\cal{L}}_{eff}^{c + (e + 2) \alpha} \ , \label{eq: masssize}
\end{eqnarray}
with\,:

\begin{displaymath}
{\cal{A}}(\alpha) = 10^{a + b\alpha} \ .
\end{displaymath}
Using Eqs.(\ref{eq: jz}) and (\ref{eq: defleff}), it is trivial to
show that\,:

\begin{eqnarray}
{\cal{L}}_{eff}^{c + (e + 2) \alpha} & = & \left \{ \frac{{\rm
e}^{b_n} b_n^{n(1 - p_n)} \Gamma(2n)}{2 \Gamma[n(3 - p_n)]}
\right \}^{\frac{c + (e + 2) \alpha}{2}} \nonumber \\
~ & \times & \left ( \frac{G \Upsilon_{\star} I_e}{L_0^2} \right
)^{\frac{c + (e + 2) \alpha}{2}} \ R_{eff}^{\frac{3[c + (e + 2)
\alpha]}{2}} \ . \label{eq: leffpow}
\end{eqnarray}
Inserting this relation into Eq.(\ref{eq: masssize}) and using
logarithmic units, it is then only a matter of algebra to finally
get\,:

\begin{equation}
\log{R_{eff}} = \log{R_{10}} + \gamma \log{\left (
\frac{M_{\star}}{2 {\times} 10^{10} \ h^{-1} \ {\rm M_{\odot}}}
\right )} \label{eq: sizemass}
\end{equation}
with\,:

\begin{equation}
\gamma = \left [ \frac{(3e + 6) \alpha + (3c + 4)}{2} \right
]^{-1} \ , \label{eq: defalpha}
\end{equation}

\begin{equation}
\log{R_{10}} = \gamma \log{(2 {\times} 10^{10} h^{-1} M_{\odot})}
- \log{{\cal{N}}(\alpha)} \ , \label{eq: defr10}
\end{equation}

\begin{eqnarray}
{\cal{N}}(\alpha) & = & \frac{2^{(6\alpha + 13)/2} \pi^{\alpha +
4} \Gamma(\alpha + 1)}{\Gamma(\alpha + 3/2)} \left \{ \frac{n
\Gamma[n(2 - p_n)]}{b_n^{n(2 - p_n)}} \right \}^{\alpha + 3/2}
\nonumber \\ ~ & \times & \left \{ \frac{{\rm e}^{b_n} b_n^{n(1 -
p_n)} \Gamma(2n)}{\Gamma[n(3 - p_n)]} \right \}^{\frac{(e +
2)\alpha + (c + 2)}{2}} \Upsilon_\star I_e {\cal{A}}(\alpha)
\nonumber \\ ~ & \times & \left ( \frac{G \Upsilon_{\star} I_e}{2
L_0^2} \right )^{\frac{(e + 2)\alpha + c}{2}} \ \eta_a^{f \alpha +
d} \ . \label{eq: defnorm}
\end{eqnarray}
Should all the ETGs have the same values of the photometric
parameters $(n, I_e)$, the stellar $M/L$ ratio $\Upsilon_{\star}$
and the anisotropy parameters $(\alpha, \eta_a)$, then their sizes
and masses should turn out to be perfectly correlated as predicted
by Eq.(\ref{eq: sizemass}). Observationally, such a correlation is
indeed found, but with an intrinsic scatter which may be obviously
due to the scatter in the quantities entering the definitions of
$\gamma$ and $R_{10}$. Assuming, however, as a first approximation
that $n = 4$ for all galaxies (i.e., all the ETGs follow the de
Vaucouleurs surface brightness profile) and that $(\alpha,
\eta_a)$ do not change on a case\,-\,by\,-\,case basis, we can
then estimate $\alpha$ and $L_0$ from the observed values of
$\gamma$ and $R_{10}$ once the fitting coefficients $(a, b, c, d,
e, f)$ have been determined. This latter task can be easily done
by computing the mass integral over a grid in the $(\alpha,
{\cal{L}}_{eff}, \eta_a)$ space for a fixed value of $n$. In
particular, for $n = 4$, we find\,:

\begin{displaymath}
(a, b, c, d, e, f) \simeq (-2.357, 1.780, -0.017, 0.223, -2.316,
0.114)
\end{displaymath}
fits very well the mass integral in Eq.(\ref{eq: defmassint}) with
a mean percentage residual $\langle \Delta {\cal{I}}/{\cal{I}}
\rangle \simeq 0.1\%$ and $(\Delta {\cal{I}}/{\cal{I}})_{rms}
\simeq 6\%$. With these values, the slope of the size\,-\,mass
relation turns out to be\,:

\begin{displaymath}
\gamma \simeq (1.974 - 0.474 \ \times \ \alpha)^{-1}
\end{displaymath}
so that, setting $\gamma = 0.56$ as observationally found by Shen
et al. (2003) for ETGs with $M_{\star} > 2 {\times} 10^{10} \
h^{-1} \ {\rm M_{\odot}}$, we finally get $\alpha \simeq 0.40$,
i.e. the velocity anisotropy is tangential for $r/r_a <
\sqrt{\alpha} \simeq 0.63$ and radial elsewhere. For lower mass
systems, Shen et al. (2003) find $\gamma = 0.14$ thus giving
$\alpha = -10.2$. Such a large negative value should argue in
favor of a fully radial anisotropy, but also imply a quite strong
angular momentum term in the anisotropic DF which seems quite
unusual. Moreover, we have computed the mass integral only for
$\alpha > 0$ so that our approximation for $\log{{\cal{I}}}$
should not be extrapolated to models with negative $\alpha$.

It is worth stressing, however, that the surface brightness
profile of dwarf ellipticals is better fitted by Sersic profiles
with $n = 1$ rather than $n = 4$ so that the above relation
between $\alpha$ and $\gamma$ does not apply anymore. We have
therefore recomputed the mass integral over the same grid, but
setting now $n = 1$. This gives us\,:

\begin{displaymath}
(a, b, c, d, e, f) \simeq (-1.641, 0.644, -0.927, -0.463, -2.491,
0.662)
\end{displaymath}
thus giving\,:

\begin{displaymath}
\gamma \simeq (0.609 - 0.736 \ \times \ \alpha)^{-1} \ .
\end{displaymath}
Unfortunately, setting $\gamma = 0.14$ and solving for $\alpha$,
we get $\alpha \simeq -8.9$ so that we are unable to escape the
same problems as with $n = 4$. Moreover, our approximating
expression (\ref{eq: ansatz}) now works well only over a limited
range in $({\cal{L}}_{eff}, \eta_a)$ with significantly worse
residuals ($\langle \Delta {\cal{I}}/{\cal{I}} \rangle \simeq
6\%$, $(\Delta {\cal{I}}/{\cal{I}})_{rms} \simeq 14\%$). We
therefore caution the reader to not overrate the validity of the
$\gamma$\,-\,$\alpha$ relation for $n = 1$ Sersic models.

\section{A dark size\,-\,mass relation}

Although our main aim was to investigate the size\,-\,mass
relation for the visible component of ETGs, it is worth stressing
that the above procedure is quite general. As an interesting
application, we consider dark haloes since it has been proposed
\cite{Merritt} that the PS model best approximates the density
profile of dark haloes coming out from dark matter only
simulations. Actually, the PS parameters change from one case to
another so that a general rule could not be extracted.
Nevertheless, the value of $n$ is not too dispersed so that we
will repeat the above calculation setting $n$ to the average value
$n = 3.58$. As a result, we obtain for the slope $\gamma$ of the
$R_{eff}$\,-\,$M$ relation\,:

\begin{displaymath}
\gamma \simeq 1.915 - 0.263 \times \alpha \ .
\end{displaymath}
An important remark is in order here. Since we are now using the
PS model for the dark halo, $R_{eff}$ and $M$ can no more be
estimated unless one fits the model to kinematical (such as the
velocity dispersion profile) or lensing (e.g., Einstein rings)
data. As such, we cannot provide any estimate neither for $\alpha$
or $\gamma$. However, it is convenient to reparameterize the
size\,-\,mass relation for haloes in terms of the concentration
parameter and the virial mass. Without loss of precision, we can
simply identify the halo virial mass with the PS total mass. In
order to define a concentration for the PS halo, we first
introduce the scale radius $r_s$ defined by the condition\,:

\begin{displaymath}
\frac{d\ln{\rho}}{d\ln{r}} = -2
\end{displaymath}
in close analogy with the meaning of $r_s$ in the NFW model. Using
Eq.(\ref{eq: jr}), we easily get\,:

\begin{equation}
r_s = \left [ \frac{n (2 - p_n)}{b_n} \right ]^n R_{eff} \ .
\label{eq: defrs}
\end{equation}
The concentration $c_{PS}$ may be then defined as $c_{PS} =
R_{vir}/r_s$, and, using $R_{vir} \propto M_{vir}^{1/3}$ and
$R_{eff} \propto M_{vir}^{\gamma}$, we finally get\,:

\begin{displaymath}
c_{PS} \propto M_{vir}^{- \gamma_{DM}}
\end{displaymath}
with\,:

\begin{displaymath}
\gamma_{DM} = \gamma - 1/3 \simeq \frac{1.085 + 0.263 \times
\alpha}{3(1.915 - 0.263 \times \alpha)} \ .
\end{displaymath}
For $0 \le \alpha \le 1$, we get $0.19 \le \gamma_{DM} \le 0.27$,
so that the concentration is a mild decreasing function of the
virial mass. Such a behavior is in agreement with the results of
simulations, but a direct comparison is not possible because of
the use of different halo models.

\section{Conclusions}

In the present paper, we studied the effect of injecting angular
momentum in a Sersic isotropic profile. Our starting motivation
was to investigate the role of angular momentum in explaining the discrepancy
between FP projections (in our case, the $R_{eff}$\,-\,$M_{\star}$
relation) and observations. Moreover, such a work can also shed
light on the differences between dwarf and normal ellipticals as
evidenced by the same FP projections.

Injecting angular momentum makes the velocity dispersion tensor
anisotropic so that the DF has to be accordingly modified. The
Osipkov\,-\,Merritt parameterization provides a valuable tool to
infer the corresponding DF and hence the modified density profile.
As a consequence, the size\,-\,mass relation turns out to be
changed with respect to the isotropic case thus tilting the
$\log{R_{eff}}$\,-\,$\log{M_{\star}}$ relation with respect to the
isotropic value. As we have shown, the slope $\gamma$ of the
size\,-\,mass relation is an easy function of the anisotropy
parameter $\alpha$. For massive ($M_{\star} \ge 2 \times 10^{10} \
h^{-1} \ {\rm M_{\odot}}$ETGs, our model may be reconciled with
the data provided the anisotropy profile is tangential in the
inner regions (i.e., for $r/r_a < 0.6$) to become then radial
elsewhere. It is worth noting that, while $\alpha$ is the only
parameter controlling the slope of the size\,-\,mass relation, all
the three quantities $(\alpha, r_a, L_0)$ enter in determining the
zeropoint of the correlation. As a consequence, while $\alpha$ can
be determined by $\gamma$, we cannot infer a unique value for the
anisotropy scale $r_a/R_{eff}$ unless we estimate somewhat the
reference angular momentum $L_0$.

Actually, the unique way to break this degeneracy relies on
fitting the velocity dispersion profile for a statistically
meaningful sample of ETGs in order to determine, on a
case\,-\,by\,-\,case basis, the value of $(\alpha, r_a/R_{eff})$.
Such a test is also mandatory in order to check the validity of
the assumed OM parameterization of the anisotropy profile.
Moreover, the distribution of the $(\alpha, r_a/R_{eff})$ values
can provide an estimate of the expected scatter in the
$\log{R_{eff}}$\,-\,$\log{M_{\star}}$ relation thus giving a
further cross check of the model.

It is worth wondering whether our results can help in elucidating
why dwarfs and giant ellipticals present surprising structural
differences. Indeed, while dEs appear as the natural low luminosity
counterpart of giant Es, they do not follow the same scaling relations
as the latter ones. For instance, the $R_{eff}$\,-\,$M_{\star}$ relation 
is rather shallow with a slope $\gamma = 0.14$ significantly smaller
than the one $(\gamma = 0.56)$ of the high mass ellipticals. To this end, 
we first remember that, while both normal and dEs formed from the gravitational
collapse of primordial density fluctuations, their structural 
and stellar population properties depend on the ability of the baryons in a 
given overdense region to cool and form stars. In particular, a feedback mechanism
that transfers energy back to the interstellar medium is usually invoked to 
explain the dEs structure. For instance, models that invoke
the cessation of star formation by supernova driven winds provide
a plausible explanation for the variation of density (surface
brightness) and metallicity (color) with luminosity (Larson 1974;
Saito 1979; Vader 1986; Dekel \& Silk 1986; Arimoto \& Yoshii
1987). On the other hand, in Del Popolo (2002) and Del Popolo et al. (2005), 
it has been shown that a similar role may also be played by the 
acquisition of angular momentum. Indeed, the acquisition of a higher amount 
of angular momentum by a less dense region helps contrasting the 
gravitational collapse and hence prevents the cooling of baryons 
and the star formation. 

Based on this latter model, we therefore expect thart injecting angular
momentum in the dEs DF produces a decrease of mass infall towards the centre
thus giving rise to a structure with lower central surface brightness. In order 
to qualitatively confirm this picture, it is worth remembering that dEs present roughly 
three different morphologies with most bright dwarfs $(M_B \le -16)$ presenting a distinct
luminosity spike in their centre (referred to as the nucleus). Lower luminosities dwarfs do 
not present such a nuclear region and may be divided in S0\,-\,like and elliptical\,-\,like 
with this latter tipycally having a very low surface brightness. 
The different morphology with luminosity can be explained, in the
previous scheme, as follows: bright dwarfs are born from higher
density peaks, less subject to tidal torque, and as a consequence
one expects that more mass freely fall to the centre giving rise
to the central nucleus. Low surface brightness types are born from
lower density peaks which suffer a larger tidal torque. This
produce galaxies without a central nucleus, more extended and
having low surface brightness. We have previously seen that in
order to explain the $R_{eff}$\,-\,$M_{\star}$ scaling law for dEs we need a
negative value of $\alpha$ implying a large angular momentum term
in the anisotropic DF, somehow in agreement with this qualitative picture.
Moreover, the finding that very different values for $\alpha$ are needed in
order to explain the reproduced the observed size\,-\,mass relation for 
dEs and normal Es confirms our notion that they have a
very different dynamical origin. 

While the outcome of this simplified investigation goes in the right direction, 
more work is needed in order to put these preliminary results on a firmer ground. 
As a first major improvement, one has to give off the hypothesis of spherical symmetry
allowing for rotational flattening. This is particularly important since some evidences 
argue in favour of dEs being the final evolutionary state of tidal dwarfs galaxies 
(Gentile et al. 2007) thus originating from rotationally supported gas rich tidal arms. 
It is therefore worth investigating whether injecting angular momentum in the DF of a 
flattened PS model may allow us to still preserve our predicted $R_{eff}$\,-\,$M_{\star}$ 
relation and, at the same time, solving the problem of the negative $\alpha$ needed to 
reproduce the observed slope. Should this be the case, we could have a strong evidence 
in favour of angular momentum playing the leading role in determining the 
structure of dwarf ellipticals.

\section*{Acknowledgements}

It is a pleasure to thank an anonymous referee for his/her comments which have 
helped to significantly ameliorate the presentation. VFC is supported by Regione 
Piemonte and Universit\`a di Torino. Partial support from INFN project PD51 is 
acknowledged too.

\end{document}